\font\ottorm=cmr8
\headline{\hss \ottorm Draft \#1}

\newcount\mgnf\newcount\tipi\newcount\tipoformule
\newcount\aux\newcount\piepagina\newcount\xdata
\mgnf=0
\aux=0           
\tipoformule=1   
\piepagina=2     
\xdata=1         
\def\Di{31\ dicembre\ 1997}

\ifnum\mgnf=1 \aux=0 \tipoformule =1 \piepagina=1 \xdata=1\fi
\newcount\bibl
\ifnum\mgnf=0\bibl=0\else\bibl=1\fi
\bibl=0

%
%
%
%
\ifnum\bibl=0
\def\ref#1#2#3{[#1#2]\write8{#1@#2}}
\def\rif#1#2#3#4{\item{[#1#2]} #3}
\fi

\ifnum\bibl=1
\openout8=ref.b
\def\ref#1#2#3{[#3]\write8{#1@#2}}
\def\rif#1#2#3#4{}

\fi

\def\9#1{\ifnum\aux=1#1\else\relax\fi}
\ifnum\piepagina=0 \footline={\rlap{\hbox{\copy200}\
$\st[\number\pageno]$}\hss\tenrm \foglio\hss}\fi \ifnum\piepagina=1
\footline={\rlap{\hbox{\copy200}} \hss\tenrm \folio\hss}\fi
\ifnum\piepagina=2\footline{\hss\tenrm\folio\hss}\fi

\ifnum\mgnf=0 \magnification=\magstep0
\hsize=12.8truecm\vsize=22.5truecm \parindent=4.pt\fi
\ifnum\mgnf=1 \magnification=\magstep1
\hsize=16.0truecm\vsize=22.5truecm\baselineskip14pt\vglue5.0truecm
\overfullrule=0pt \parindent=4.pt\fi

\let\a=\alpha\let\b=\beta \let\g=\gamma \let\d=\delta
\let\e=\varepsilon \let\z=\zeta \let\h=\eta
\let\k=\kappa \let\l=\lambda  \let\n=\nu
\let\x=\xi \let\p=\pi \let\r=\rho \let\s=\sigma \let\t=\tau
 \let\f=\varphi  \let\o=\omega
 \let\G=\Gamma \let\D=\Delta 
   \let\F=\Phi
 \let\O=\Omega 
{\count255=\time\divide\count255 by 60 \xdef\oramin{\number\count255}
\multiply\count255 by-60\advance\count255 by\time
\xdef\oramin{\oramin:\ifnum\count255<10 0\fi\the\count255}}
\def\ora{\oramin }

\ifnum\xdata=0
\def\data{\number\day/\ifcase\month\or gennaio \or
febbraio \or marzo \or aprile \or maggio \or giugno \or luglio \or
agosto \or settembre \or ottobre \or novembre \or dicembre
\fi/\number\year;\ \ora}
\def\Di{\number\day\ \ifcase\month\or gennaio \or
febbraio \or marzo \or aprile \or maggio \or giugno \or luglio \or
agosto \or settembre \or ottobre \or novembre \or dicembre
\fi\number\year}
\else
\def\data{\Di}
\fi

\setbox200\hbox{$\scriptscriptstyle \data $}
\newcount\pgn \pgn=1
\def\foglio{\number\numsec:\number\pgn
\global\advance\pgn by 1} \def\foglioa{A\number\numsec:\number\pgn
\global\advance\pgn by 1}
\global\newcount\numsec\global\newcount\numfor \global\newcount\numfig
\gdef\profonditastruttura{\dp\strutbox}
\def\senondefinito#1{\expandafter\ifx\csname#1\endcsname\relax}
\def\SIA #1,#2,#3 {\senondefinito{#1#2} \expandafter\xdef\csname
#1#2\endcsname{#3} \else \write16{???? ma #1,#2 e' gia' stato definito
!!!!} \fi} \def\etichetta(#1){(\veroparagrafo.\veraformula) \SIA
e,#1,(\veroparagrafo.\veraformula) \global\advance\numfor by 1
\9{\write15{\string\FU (#1){\equ(#1)}}} \9{ \write16{ EQ \equ(#1) == #1
}}} \def \FU(#1)#2{\SIA fu,#1,#2 }
\def\etichettaa(#1){(A\veroparagrafo.\veraformula) \SIA
e,#1,(A\veroparagrafo.\veraformula) \global\advance\numfor by 1
\9{\write15{\string\FU (#1){\equ(#1)}}} \9{ \write16{ EQ \equ(#1) == #1
}}} \def\getichetta(#1){Fig.  \verafigura \SIA e,#1,{\verafigura}
\global\advance\numfig by 1 \9{\write15{\string\FU (#1){\equ(#1)}}} \9{
\write16{ Fig.  \equ(#1) ha simbolo #1 }}} \newdimen\gwidth \def\BOZZA{
\def\alato(##1){ {\vtop to \profonditastruttura{\baselineskip
\profonditastruttura\vss
\rlap{\kern-\hsize\kern-1.2truecm{$\scriptstyle##1$}}}}}
\def\galato(##1){ \gwidth=\hsize \divide\gwidth by 2 {\vtop to
\profonditastruttura{\baselineskip \profonditastruttura\vss
\rlap{\kern-\gwidth\kern-1.2truecm{$\scriptstyle##1$}}}}} }
\def\alato(#1){} \def\galato(#1){}
\def\veroparagrafo{\number\numsec}\def\veraformula{\number\numfor}
\def\verafigura{\number\numfig}
\def\geq(#1){\getichetta(#1)\galato(#1)}
\def\Eq(#1){\eqno{\etichetta(#1)\alato(#1)}}
\def\eq(#1){\etichetta(#1)\alato(#1)}
\def\Eqa(#1){\eqno{\etichettaa(#1)\alato(#1)}}
\def\eqa(#1){\etichettaa(#1)\alato(#1)}
\def\eqv(#1){\senondefinito{fu#1}$\clubsuit$#1\write16{No translation
for #1} \else\csname fu#1\endcsname\fi}
\def\equ(#1){\senondefinito{e#1}\eqv(#1)\else\csname e#1\endcsname\fi}
\openin13=#1.aux \ifeof13 \relax \else \input #1.aux \closein13\fi
\openin14=\jobname.aux \ifeof14 \relax \else \input \jobname.aux
\closein14 \fi \9{\openout15=\jobname.aux} \newskip\ttglue


\font\titolo=cmbx10 scaled \magstep1
\font\ottorm=cmr8\font\ottoi=cmmi7\font\ottosy=cmsy7
\font\ottobf=cmbx7\font\ottott=cmtt8\font\ottosl=cmsl8\font\ottoit=cmti7
\font\sixrm=cmr6\font\sixbf=cmbx7\font\sixi=cmmi7\font\sixsy=cmsy7

\font\fiverm=cmr5\font\fivesy=cmsy5\font\fivei=cmmi5\font\fivebf=cmbx5
\def\ottopunti{\def\rm{\fam0\ottorm}\textfont0=\ottorm%
\scriptfont0=\sixrm\scriptscriptfont0=\fiverm\textfont1=\ottoi%
\scriptfont1=\sixi\scriptscriptfont1=\fivei\textfont2=\ottosy%
\scriptfont2=\sixsy\scriptscriptfont2=\fivesy\textfont3=\tenex%
\scriptfont3=\tenex\scriptscriptfont3=\tenex\textfont\itfam=\ottoit%
\def\it{\fam\itfam\ottoit}\textfont\slfam=\ottosl%
\def\sl{\fam\slfam\ottosl}\textfont\ttfam=\ottott%
\def\tt{\fam\ttfam\ottott}\textfont\bffam=\ottobf%
\scriptfont\bffam=\sixbf\scriptscriptfont\bffam=\fivebf%
\def\bf{\fam\bffam\ottobf}\tt\ttglue=.5em plus.25em minus.15em%
\setbox\strutbox=\hbox{\vrule height7pt depth2pt width0pt}%
\normalbaselineskip=9pt\let\sc=\sixrm\normalbaselines\rm}
\catcode`@=11
\def\footnote#1{\edef\@sf{\spacefactor\the\spacefactor}#1\@sf
\insert\footins\bgroup\ottopunti\interlinepenalty100\let\par=\endgraf
\leftskip=0pt \rightskip=0pt \splittopskip=10pt plus 1pt minus 1pt
\floatingpenalty=20000
\smallskip\item{#1}\bgroup\strut\aftergroup\@foot\let\next}
\skip\footins=12pt plus 2pt minus 4pt\dimen\footins=30pc\catcode`@=12
\newdimen\xshift \newdimen\xwidth \newdimen\yshift
\def\ins#1#2#3{\vbox to0pt{\kern-#2 \hbox{\kern#1
#3}\vss}\nointerlineskip} \def\eqfig#1#2#3#4#5{ \par\xwidth=#1
\xshift=\hsize \advance\xshift by-\xwidth \divide\xshift by 2
\yshift=#2 \divide\yshift by 2 \line{\hglue\xshift \vbox to #2{\vfil #3
\includegraphics{#4.ps} }\hfill\raise\yshift\hbox{#5}}} \def\8{\write13}

\def\V#1{{\,\underline#1\,}}
\def\T#1{#1\kern-4pt\lower9pt\hbox{$\widetilde{}$}\kern4pt{}}
\let\dpr=\partial \let\io=\infty
\def\fra#1#2{{#1\over#2}}\let\0=\noindent
\def\guida{\leaders\hbox to 1em{\hss.\hss}\hfill}
\def\tende#1{\vtop{\ialign{##\crcr\rightarrowfill\crcr
\noalign{\kern-1pt\nointerlineskip} \hglue3.pt${\scriptstyle
#1}$\hglue3.pt\crcr}}} \def\otto{\
{\kern-1.truept\leftarrow\kern-5.truept\to\kern-1.truept}\ }

\def\pagina{\vfill\eject}\let\ciao=\bye

\def\st{\scriptscriptstyle}
\def\*{\vskip0.3truecm}

\def\lis#1{{\overline #1}}\def\eg{\hbox{\it e.g.\ }}
\def\ap{\hbox{\it a priori\ }}
\def\ie{\hbox{\it i.e.\ }}

\def\fiat{{}}
\def\\{\hfill\break} \def\={{ \; \equiv \; }}
\def\Im{{\rm\,Im\,}}\def\Re{{\rm\,Re\,}}

\ifnum\aux=1\BOZZA\else\relax\fi
\ifnum\tipoformule=1\let\Eq=\eqno\def\eq{}\let\Eqa=\eqno\def\eqa{}
\def\equ{{}}\fi
\def\defi{\,{\buildrel def \over =}\,}

\def\pallino{{\0$\bullet\;$}}\def\1{\ifnum\mgnf=0\pagina\else\relax\fi}
\def\W#1{#1_{\kern-3pt\lower6.6truept\hbox to 1.1truemm
{$\widetilde{}$\hfill}}\kern2pt\,}
\def\Re{{\rm Re}\,}\def\Im{{\rm Im}\,}\def\DD{{\cal D}}

\def\igb{
\mathop{\raise4.pt\hbox{\vrule height0.2pt depth0.2pt width6.pt}
\kern0.3pt\kern-9pt\int}}

\def\FINE{
\*
\0{\it Internet:
Author's preprints downloadable (latest version) at:

\centerline{\tt http://chimera.roma1.infn.it}
\centerline{\tt http://www.math.rutgers.edu/$\sim$giovanni}

\0Mathematical Physics Preprints (mirror) pages.\\
\sl e-mail: Giovanni.Gallavotti@roma1.infn.it,
Guido.Gentile@roma1.infn.it,\\
Vieri.Mastropietro@roma1.infn.it
}}

\def\II{{\cal I}}
\def\NN{{\cal N}}

\def\II{{\cal I}}\def\TT{{\cal T}}
\def\NN{{\cal N}}

\def\HH{{\cal H}}

\def\aa{{\V \a}}\def\nn{{\V\n}}\def\AA{{\V A}}\def\hh{{\V h}}

\def\oo{{\V \o}}\def\nn{{\V\n}}

\def\ndpr{{\kern1pt\raise 1pt\hbox{$\not$}\kern1pt\dpr\kern1pt}}
\def\Ndpr{{\kern1pt\raise 1pt\hbox{$\not$}\kern0.3pt\dpr\kern1pt}}
\def\hdp{{\h^{\fra12}}}\def\hdm{{\h^{-\fra12}}}
\def\CS{{\cal S}}

\font\cs=cmcsc10


\fiat
\centerline{\titolo Hamilton-Jacobi equation,
heteroclinic chains and}
\centerline{\titolo Arnol'd diffusion in three time scales systems}
\*\*

\centerline{\bf G. Gallavotti, G. Gentile, V. Mastropietro}
\*
\centerline{Universit\`a di Roma 1,2,3 }
\centerline{\Di}
\*\*\*
\font\cs=cmcsc10

\line{\vtop{
\line{\hskip1.5truecm\vbox{\advance \hsize by -3.1 truecm
\\{\cs Abstract.}
{\it Interacting systems consisting of two rotators and a point mass
near a hyperbolic fixed point are considered, in a case in which the
uncoupled systems have three very different characteristic time
scales.  The abundance of quasi periodic motions in phase space is
studied via the Hamilton--Jacobi equation.  The main result, a high
density theorem of invariant tori, is derived by the classical
canonical transformation method extending previous results.  As an
application the existence of long heteroclinic chains (and of Arnol'd
diffusion) is proved for systems interacting through a trigonometric
polynomial in the angle variables.}}  \hfill} }}
\*
\0{\it Keywords:
\sl Hamilton Jacobi, KAM, Arnold diffusion, homoclinic splitting}
\*\*
\0{\bf\S1. The system.}
\numsec=1\numfor=1\*

Let $(\a_1,\a_2)=\aa\in {\bf T}^2$ be a pair of angles;
let $\AA=(A_1,A_2)\in {\bf R}^2$ be their conjugate momenta
or ``{\it actions}'', and let $(p,q)\in {\bf R}^2$ be a
further pair of canonically conjugate coordinates. We consider the
Hamiltonian function, depending on two dimensionless parameters,
$\e,\h>0$, defined by:
$$\HH=h(\h^{\fra12} A_1)+\hdm{\o_2A_2}+
G(p q,\h^{\fra12}A_1)+\e f(\h^{\fra12}A_1,\aa,p,q)\Eq(1.1)$$
We suppose that $h,G,f$ are real analytic functions of their arguments in
the domain $\h^{\fra12}|A_1|\le R$, $|p|,|q|\le r, \aa\in {\bf T}^2$ for
some $R,r>0$. We {\it also suppose} that $f$ is a trigonometric polynomial
of degree $N$ in the variables $\aa$.

For $\e=0$ the Hamiltonian \equ(1.1) represents, physically,
a system consisting of two rotators and a point mass near
an unstable fixed point $(p=q=0)$.

The instability time scale of the unstable equilibrium is $g$ $\=$
$g(pq,\hdp A_1)$ $\defi$ $\dpr_xG(x,$ $\hdp A_1)|_{x=pq}$ $=$ $O(1)$
if, as we shall suppose, $g>0$.

Define $\hdp\o_1$ $\=$
$\hdp\o_1(pq,\hdp A_1)$ $=$ $\dpr_{A_1}[h(\hdp A_1)$ $+$
$G(p q,\h^{\fra12}A_1)]$:
the two rotators rotate with angular velocities $\hdp\o_1$ and
$\hdm\o_2$ with ratio $O(\h)$ if, as we shall suppose, $0\le\tilde
\o_a$ $\le$ $|\o_1|$ $\le$ $\tilde \o_b$
for suitable constants $\tilde\o_a,\tilde\o_b$.

Hence the system has three time scales of respective orders
$\hdm,1,\hdp$ called respectively: { \it slow} (rotator \#1), {\it
normal} (unstable motion), {\it fast} (rotator \#2). We suppose $\h<1$
so that the fast rotator really is the {\it isochronous} one and the
slow rotator is the {\it anisochronous} one.
\*

\0{\bf\S2. Characteristic parameters.}
\numsec=2\numfor=1\*

A few {\it characteristic parameters} can be associated with the
system because of the analyticity assumptions.

Let $2\r_0,2\x_0,(2\k_0)^{\fra12},(2\k_0)^{\fra12},2\k_0$ measure the
size in the complex planes of a holomorphy domain in the
$A_1,\a_i,p,q$ and $x=p q$ variables, respectively. We denote:
$$\eqalign{
Q_{\r}=&\big\{ |\Re A_1|\le  \hdm (R+\r),\, |\Im A_1|\le \hdm \r\big\}\cr
U_{\x}=&\big\{|\Im \a_j|\le \x\big\},\qquad V_{\k}=
\big\{|p|,|q|\le \k^{\fra12}\big\},\qquad S_{\k}=\big\{
|x|\le \k\big\}\cr}\Eq(2.1)$$
Our analyticity hypotheses imply the existence of $\r_0,\x_0,\k_0>0$
such that $h(\hdp A_1)$ is holomorphic in $Q_{2\r_0}$, $G(x,\hdp A_1)$
is holomorphic in $S_{2\k_0}\times Q_{2\r_0}$ and $f$ in
$Q_{2\r_0}\times U_{2\x_0}\times V_{2\k_0}\defi
\DD_{2\r_0,2\x_0,2\k_0}$.

Suppose:\\
(i) $g(pq,\hdp A_1) \defi \dpr_x G(x,\hdp A_1)|_{x=pq} \ne0$;\\
(ii) $\o_1(pq,\hdp A_1)\defi\dpr h(\hdp A_1)+\dpr G(pq,\hdp
A_1)\ne0$, where $\dpr$ denotes differentiation with respect to the
argument $a=\hdp A_1$;\\
(iii) $m_1(pq,\hdp A_1)\defi \dpr^2 h(\hdp A_1)+\dpr^2 G(pq,\hdp A_1)
\ne0$.

Hence we can define $\G_0,E_0, M_0>0$ so that:
$$\eqalign{
&0<\l_0=\min\{\r_0,\k_0\},\qquad
|f|,|h|,|G|< E_0 \cr
&0<\G_0 < |\o_1(pq,\hdp A_1)|,
\,|\o_2|, |g(pq,\hdp A_1)|< E_0\l^{-1}_0\cr
&  0<M_0^{-1} < |m_1(pq,\hdp A_1)| \cr}\Eq(2.2)$$
in the holomorphy domain $\DD_{2\r_0,2\x_0,2\k_0}$. We suppose $\h<1$
(see \S1) and, for simplicity, $\x_0,\e<1$.

The parameters $E_0,\G_0,M_0,\l_0,\x_0$ will be called {\it
characteristic parameters} of the system (note that $\G_0\l_0,
\tilde\o_b\l_0<E_0$).
{\it Only} the dimensionless parameters $\x_0$,
$E_0\G_0^{-1}\l_0^{-1}$ and $M_0E_0\l_0^{-2}$ will control the
constants in the main bounds (see \equ(4.3) and \equ(5.5)).

\*
\0{\bf\S3. Hamilton--Jacobi equation.}
\numsec=3\numfor=1\*

The Hamilton--Jacobi equation for the Hamiltonian \equ(1.1)
will be an equation with two unknown functions
$\F(\hdp A'_1,\aa,p',q)$ and $\tilde G(x, \hdp A'_1)$ given by:
$$\eqalign{
&h(\hdp(A'_1+\dpr_{\a_1}\F))+\hdm
\o_2(A'_2+\dpr_{\a_2}\F)+G((p'+\dpr_q\F) q,
\hdp(A'_1+\dpr_{\a_1}\F))\cr &
\ +\e f(\hdp(A'_1+\dpr_{\a_1}\F),\aa,p'+\dpr_q\F,q)=\cr
&=h(\hdp A'_1)+\hdm\o_2 A'_2 +G(p'(q+\dpr_{p'}\F),\hdp A_1')
+\tilde G(p' (q+\dpr_{p'}\F),\hdp A_1')\cr}\Eq(3.1)$$

If the functions $\F,\tilde G$ existed one could set:
$$\eqalign{\AA=&\AA'+\dpr_\aa\F,\qquad p=p'+\dpr_q \F\cr
\aa'=&\aa+\dpr_{\AA'}\F,\qquad q'=q+\dpr_{p'}\F\cr}\Eq(3.2)$$
and the change of variables $(\AA,\aa,p,q)\otto(\AA',\aa',p',q')$,
{\it where defined}, would cast the Hamiltonian \equ(1.1) into the form:
$$\HH_0 =h(\hdp A'_1)+\hdm \o_2 A'_2+G(p' q',\hdp A'_1)+
\tilde G(p' q',\hdp A'_1)\Eq(3.3)$$
thereby integrating the system, because the motions in the new coordinates
would become $\AA'={\rm const.}$, $\aa'\to\aa'+\oo' t$, $p'\to p'
e^{-\lis g 't}$ and $q'\to q'e^{\lis g 't}$ with:
$$\eqalignno{
\lis g' & \=\lis g'
(x',\hdp A'_1)= \dpr_{x'} G(x',\hdp A'_1)+\dpr_{x'}\tilde G(x',
\hdp A'_1) & \eq(3.4) \cr \oo' & \=\oo'(x',\hdp A'_1)
=\big(\hdp\dpr' h(\hdp A'_1)+\hdp\dpr' G(x',\hdp A'_1)+\hdp\dpr'
\tilde G(x',\hdp A'_1),\hdm \o_2\big)\cr} $$
where $x'=p'q'$ and $\dpr'$ denotes differentiation with respect
to the argument $a'=\hdp A'_1$.

This is in fact, as it is well known, {\it not possible} in general.
However one proves the following result.

\*
\0{\cs Proposition.}
{\it There exists a canonical change of variables
like \equ(3.2) such that\\
(a) it is of class $C^\io$ in all variables and in $\e$ for $\e$ small
enough;\\
(b) it is close to the identity within $O(\e)$ in $C^\io$;\\
(c) there is a function $\tilde G(x',\hdp A_1')$ which is
of class $C^\io$ in
all variables, including $\e$, which  is divisible by $\e$ and such that
in the new variables the Hamiltonian coincides
together with its derivatives, with \equ(3.3), on the set (without
interior points) of the $(A'_1,p,'q',\aa')$'s such that
the rotation vector $\oo'\=\oo'(p'q',\hdp A'_1)$ defined in \equ(3.4)
verifies the {\it Diophantine condition}:
$$|\oo'\cdot\nn|> C(\h)|\nn|^{-\t} \, \qquad
\hbox{for all }\V0\ne\nn\in {\bf Z}^2 \Eq(3.5)$$
with $\t\ge1$ prefixed and $C(\h)>0$, {\it provided} $\e$ is small
enough, depending on $C(\h)$ and the other parameters of the model.}

\*

The above proposition is essentially proven in [CG], \S5, Lemma 1',
(see [G2] and [P] for cases without gaps).
The proof in [CG] deals really with the subset of the set the
variables $\AA'$ and $x'=p'q'$ where $\oo'(p'q',\hdp A'_1)$ $=$
$(1+\g(p'q',\hdp A'_1))$ $\oo'(0,\hdp A'_1)$ for some $\g(x',a')$ of
class $C^\io$ in its arguments: by examining the proof one sees that
the result holds under the above more general condition.

\*

\0{\it Remarks}.
(1) Calling $W=W_{\e,C(\h)}$ the set in the phase space
such that the corresponding rotation vectors $\oo'$ verify
\equ(3.5), one says that
the Hamilton--Jacobi equation is soluble on $W$ and, in $W$, casts the
Hamiltonian \equ(1.1) into the {\it normal form} given by the
r.h.s. of \equ(3.3).

(2) Since the function $\oo'$ has non zero gradient with respect to
$\AA'$ (under the mentioned condition of smallness of $\e$; see
\equ(2.2)), the volume $W$ of phase space where \equ(3.5) holds has a
complement with measure bounded proportionally to $C(\h)$.  And at
fixed $x'=p'q'$ the measure of the subset of the interval $[-\hdm
R,\hdm R]$ on the $A'_1$ axis has also measure bounded proportionally
to $C(\h)$. The latter subset consists of a sequence of small
intervals (whose total length has size $<O(C(\h))$) that are
called ``{\it gaps}'' for natural reasons.

(3) The values of $A'_1$ such that \equ(3.5) holds with
$x'=p'q'=0$ are points where the Hamiltonian coincides with one like
\equ(3.3) {\it together with its first derivatives}. Thus they clearly
correspond to invariant tori (if $p'=q'=0$, also called ``persistent
tori'') and to their stable and unstable manifolds (if $q'=0$ or
$p'=0$, respectively, sometimes called {\it whiskers}). On such
manifolds the motion takes place following the evolution described
after \equ(3.3).

\*

If we simply apply the results in [CG], the bounds on the gaps
between the invariant tori (which persist notwithstanding the
perturbation) are found to be ``too large'' with respect to the ``{\it
homoclinic splitting}'' size (it has become usual to say that the
splitting is ``exponentially small in $\hdm$'' as $\h\to0$). The
latter, see below (remark 3 in \S5),
is the quantity of main interest in the theory of
drift and diffusion. A simple way to use this result to discuss
diffusion in phase space is to choose $\e$ very small: but it turns
out that it has to be chosen too small for the purpose of several
interesting applications (and one says that it has to be chosen
``exponentially small with $\hdm$'');
we postpone to \S5 a more technical discussion.

\*

This paper is dedicated to overcome the limitations that have been
briefly alluded to and to show that the restrictions to be put on the
size of $\e$ can be {\it greatly improved} with respect to the ones
that follow directly from previous work, (\S5 and Appendix A10 of
[CG]). Therefore we shall proceed in a somewhat different way with
respect to the one outlined above, and we shall perform a preliminary
step (\S4) before applying the analysis of [CG].  As a consequence we
shall bound from above (\S5) the spacing between invariant tori, \ie
the gaps sizes, and apply the results to show that the homoclinic
angles are (generically) {\it very large} compared to the
(dimensionless) spacing: this will imply (\S6) the existence of long
heteroclinic chains for a system consisting of a pendulum interacting
with two rotators {\it thereby deducing (by using \S 8 of [CG]) the
existence of drift and diffusion in the system} when the perturbation
parameter $\e$ is {\it smaller than a power of $\h$, rather than
exponentially small in $\hdm$}.  Finally we show (\S7) that the
perturbation can have a large component which, provided it depends
only on the fast rotator angle (or more generally provided it is
``monochromatic'', see \S7), does not alter some of the results.

\*

A problem with three time scales naturally arises in celestial
mechanics: in the theory of the precession motion of a satellite axis
due to its non sphericity. Under certain approximations, see [CG], it
is described by a Hamiltonian of the form \equ(1.1) plus a large
``monochromatic'' term. In [CG], \S12, it was shown that Arnol'd
diffusion occurs, but a preceding error on the lower bound of a
homoclinic splitting infirms the conclusions. In [GGM1] the correct
lower bound is given and combining [GGM1] with the new results of \S7 one
sees that the analysis in [CG] applies, provided that the monochromatic
term is absent (this means that the normal form method described in
\S8 of [CG] applies to the heteroclinic chains that, below, we show to exist).
Finally, in the last section, we outline a {\it possible} strategy on
how to approach and complete the analysis of the full problem of \S12
of [CG].

\*

The non standard part of this work relies strongly on the explicit
formula for the homoclinic splitting derived in [GGM1]
and called the ``{\it large angles theorem}''.

\*
\0{\bf\S4. Averaging.}
\numsec=4\numfor=1\*

We shall first perform a canonical transformation, that we shall
denote $\F_{\NN}$, to ``reduce'' the perturbation to order
$O(\e^{\NN})$, with $\NN$ an integer suitably large, but such that no
``small divisors'' appear in the definition of $\F_{\NN}$. This is
easily achieved by requiring $\NN\le\z\h^{-1}N^{-1}$,
with $\z=|\o_2|(2\tilde\o_b)^{-1}$, (note that
the existence and the relative sizes of three time scales is essential
here).  Subsequently we shall apply the results of [CG] to the
Hamiltonian expressed in the new variables; in this way we shall be
able to prove a more refined bound on the size of the gaps between
invariant tori, (and show that it is smaller than the homoclinic
splitting without the need to require $\e$ to be exponentially small
in $\hdm$).

Fixed an integer number $\NN$, we look for a function $\F_\NN(\hdp
A'_1,\aa,p',q)$ generating a canonical transformation which casts
the Hamiltonian \equ(1.1) into the form \equ(3.3) up to order $\NN$.
This means that, in the new variables, the Hamiltonian will have the
form \equ(3.3), but with a different function $\tilde G$ (that we
shall denote $\tilde G_\NN$) and {\it an extra additive term} (the new
perturbation) which will not be of order $O(\e)$ but of order
$O(\e^{\NN})$.

The functions $\F_\NN,\tilde G_\NN$ will be polynomials in $\e$:
$$\eqalign{
\F_\NN=&\e\F^{(1)}+\e^2\F^{(2)}+\ldots+\e^\NN\F^{(\NN)} \cr
\tilde G_\NN=&\e\tilde G^{(1)}+\e^2\tilde G^{(2)}+\ldots+\e^\NN\tilde
G^{(\NN)} \cr}\Eq(4.1)$$

We proceed to determine recursively the coefficients
for $\NN\le\fra12\h^{-1} N^{-1}$
in order to prove the following {\it averaging theorem}.
\*

\0{\cs Theorem 1.} {\it Fixed $\NN\le \z\h^{-1}N^{-1}$,
with $\z=|\o_2|(2\tilde\o_b)^{-1}$, the
functions $\F^{(j)},\tilde G^{(j)}$ exist for all $1\le j\le\NN$ and
are holomorphic in the ``decreasing'' domains $\DD_{\r_j,\x_j,\k_j}$
with $\x_j=\x_0-j\d$, $\r_j=\r_0 e^{-j\d}$, $\k_j=\k_0 e^{-j\d}$ for
$\d>0$ such that $\NN\d<\fra12 \x_0$, and they verify the bounds:

$$|\F^{(j)}| < D B^{j-1} (j-1)!,\qquad |\tilde G^{(j)}|< E_0 B^{j-1}
(j-1)!,\qquad j\le \NN\Eq(4.2)$$
where $D=bE_0 \G_0^{-1} D_0$ and $B=b\,(E_0\G_0^{-1}
\l_0^{-1}\x_0^{-1})^2D_0^2\NN^2$,
if the constant $b$ is large enough,
and $D_0=\hdm \x_0^{-1}(\NN\x_0^{-1})^2\log(\NN\x_0^{-1})$.

Furthermore:

\pallino Map \equ(3.2) with $\F=\F_\NN$
generates a canonical transformation from
$\DD_{\r_\NN,\x_\NN,\k_\NN}\supseteq
\DD_{\r_0 e^{-\x_0/2},\x_0/2,\k_0e^{-\x_0/2}}$
to the domain $\DD_{\r_0,\x_0,\k_0}$ if:

$$|\e|< \e_0=b' \,\Big( {\G_0^{}\l_0^{}\over E_0}
\Big)^2\fra1\NN \fra{\h }
{ (\NN\x_0^{-1})^6\log^2(\NN\x_0^{-1}) }\Eq(4.3) $$
with $b'$ small enough.

\pallino And in the new variables $(\AA',\aa',p',q')$ the Hamiltonian
\equ(1.1) becomes:

$$\eqalign{
&h(\hdp A'_1)+\hdm \o_2 A'_2+G(p'q',\hdp A'_1)+
\tilde G_\NN(p' q',\hdp A'_1)+\cr
&+\Big(\fra{\e}{2\e_0}\Big)^{\NN}
f_{\NN,\e}(\hdp A'_1,\aa',p',q')\cr}\Eq(4.4)$$
with $|\e^{-1}\tilde G_\NN|,|f_{\NN,\e}|<2 E_0$.}
\*

The proof follows immediately (for instance) from the arguments used
to prove Nekhoroshev's theorem in [G1,BG] and is given in Appendix A1,
for completeness, extending similar considerations in [CG].
\*

\0{\it Remarks.} (1) An interesting consequence
is obtained by choosing $\NN=\g\hdm$ with $\g>0$. In this case under a
condition $|\e|<\e_0=O(\h^{\fra92}/\log^2\h^{-1})$, see \equ(4.3),
the Hamiltonian \equ(1.1) can be put in the form \equ(4.4).

(2) Another interesting consequence (of the proof of the above theorem
given in Appendix A1) concerns the case with $\NN$ fixed independently
on $\h$ and a perturbation $f$ whose Fourier transform with respect to
$\aa$ does not vanish only for wave vectors $\nn$ multiples of a given
wave vector $\nn_0$ (``{\it monochromatic}'', or ``{\it integrable}'',
perturbation). In the latter case the condition is $|\e|<O(1)$ for all
$\NN$, prefixed; see the corollary discussed in the last paragraph of
Appendix A1, (although it would be easy to prove directly the simple
statement made there).

(3) As a quantitative estimate of the parameters of the new Hamiltonian
\equ(4.4) it follows from the proof in Appendix A1 that the parameters
$\l_0,\G_0, E_0, M_0$ can all be taken the same as those of the
initial Hamiltonian within a factor $2$: also $\h$ is the same while,
of course, $\e$ becomes $(\e(2\e_0)^{-1})^\NN$.

\*
\def\SX{{\cal S}}
\0{\bf\S5. Abundance of tori. Interpolation.}
\numsec=5\numfor=1\*

Given the Hamiltonian \equ(4.4), define the {\it free spectrum}
$\SX_0$ as the collection of the rotation vectors
$$ \oo=\oo(x',\hdp \AA')\defi
\big(\hdp\dpr' h(\hdp A'_1)+\hdp\dpr' G(x',\hdp A'_1),
\hdm \o_2\big) \Eq(5.1)$$
as $A'_1$ varies in $[-R\hdm,R\hdm]$ and $x'=0$
and $\dpr'$ denotes differentiation with respect to the argument
$a'=\hdp A'_1$. More generally we define $\SX$ to be the collection of
vectors \equ(5.1) as $A'_1\in[-R\hdm,R\hdm]$ and $|x'|<\k_0e^{-\x_0/2}$.

Let us consider the values $A'_1$ such that the rotation vector
$\oo\in \SX_0$ in \equ(5.1) satisfies at $x'=0$
the {\it dio\-phan\-tine condition}:
$$|\hdp\o_1 \n_1+\hdm\o_2 \n_2|> C(\h)|\nn|^{-\t}
\; \qquad \hbox{for all }\V0\ne\nn\in{\bf Z}^2 \Eq(5.2)$$
where $\hdp\o_1\=\hdp\o_1(x',\hdp A_1')$ is the first component of
$\oo$ in \equ(5.1), and $\t\ge1$, $C(\h)>0$ are Diophantine
constants: $\o_1$ has size of order $1$ (see \S1).

Such $A'_1$ occupy a set in $[-R\hdm,R\hdm]$ whose complement has
measure bounded above by:
$$b'' M_0 E_0 \l_0^{-1} C(\h) \Eq(5.3)$$
with a suitably large $b''$, because the measure of the set of $\o_1$'s
in the interval of variability of $\o_1$ ($[\tilde \o_a,\tilde\o_b]
\subseteq [0,E_0\l_0^{-1}]$,
see \S1) verifying \equ(5.2) has complement bounded
by $\tilde b'' E_0\l_0^{-1}\hdp C(\h))$,
for some constant $\tilde b''$,
and the derivative of the map from action to frequency is
$\h \dpr^2 h(pq,\hdp A_1)$, bounded below by $\h M_0^{-1}$ (see \equ(2.2)).
Hence any interval in $[-R\hdm,R\hdm]$ of size $\D(\h)> b''
M_0E_0\l_0^{-1}C(\h)$ will contain points verifying \equ(5.2).

If $C(\h)=\O e^{-s \hdm}$ with $s,\O>0$ (the factor $\O$ is here just
to fix the dimensions and it could be any constant with the right
dimensions,\eg $\o_2$ or $E_0/\l_0$ or other), then it follows from
the combination of theorem 1 above and lemma 1' of [CG] that for each
$\oo\in\SX_0$ verifying \equ(5.2) the Hamiltonian system
\equ(3.3), hence \equ(4.4), will have an invariant torus such that the
rotation vector $\oo'$ assumes the value
$\oo$, for $\h$ small enough and $|\e|<\e_0$.

In fact the ``effective coupling'', by applying theorem 1 with
$\NN=\g\hdm$, is of order $O((\e/2\e_0)^{\g\hdm})$ and, for $\g$
large, this is so small that even if multiplied by $[C(\h)/\O]^{-q}$ for
$q>0$ arbitrarily fixed gives a still very small result when $\h$ gets
small enough.

The stability of the tori and their stable and unstable manifolds or of
the motions in their vicinity depends on the size of the latter
product (for a suitably large $q$, see [CG], eq. (5.76), where $q=6$
is an estimate) which is small if $\g>qs(\log2)^{-1}$.

Note that if we simply applied lemma 1' of [CG] to the Hamiltonian
\equ(1.1) we would have found, as a bound on the convergence radius for
the parametric equations of the tori and of some of their
nearby motions, a quantity of order
$O(\e_0)$ times $[C(\h)/\O]^{q}$, which would have been exponentially
small in $\hdm$ if $C(\h)=\O e^{-s \hdm}$ with $\O,s>0$.

Another way of interpreting the analysis (see [CG]) is by saying that
there is a change of coordinates $(\AA,\aa,p,q)\otto(\AA',\aa',p',q')$,
defined in the vicinity of $|\Re A_1|<\hdm R$ and $|p|,|q|$ small,
which is of class $C^\io$ (in $(\AA',\aa',p',q')$ and $\e$) which
casts Hamiltonian \equ(1.1) or \equ(4.4) in the form \equ(3.3) on the
set of $(\AA',\aa',p',q')$'s for which the vector $\oo'=\oo'(p'q',\hdp
A'_1)$ defined in \equ(3.4) verifies \equ(3.5), \ie
it is equal to some $\oo\in\SX_0$ verifying \equ(5.2).

Furthermore, fixed $\AA'$ and $p',q'$ in such a set,
the coordinate transformation (of the
remaining two coordinates $\aa$) is analytic in $\e$ for
$|\e|<\e_0 O([C(\h)/\O]^{q\g^{-1}\hdp})$.

And the identity
between \equ(1.1) and \equ(3.3) with \equ(3.4) holds, on the set on
which the functions agree, {\it also between their first derivatives
with respect to the $(\AA',\aa',p',q')$ variables}. So that the sets
with $\AA'$ and $p'q'=x'$ fixed and $\oo'$ verifying \equ(3.5)
are invariant and the motion on them is
very simple, and described after \equ(3.3).

We stress again that the latter statement is a slight generalization
of the quoted results in [CG]: namely the latter really covers the
statements in the preceding paragraph for vectors $\oo'(x',\hdp A'_1)$
which verify \equ(3.5) {\it and} have the special form
$\oo'(x',\hdp A'_1)$ $=$ $(1+\g(x',\hdp A'_1))\,\oo'(0,\hdp A'_1)$
for some small $\g(x',\hdp A'_1)$.
This restricts us, {\it de facto}, to considering cases less general than
those that verify \equ(3.5). However, as mentioned above, if one goes
through the proof of lemmata 1,1',2 of [CG] one realizes that the
proof covers, unchanged, the more general case we quote here.

As a consequence the following {\it high density
theorem of invariant tori} also holds (corollary of
theorem 1 and of the above analysis).

\*
\0{\cs Theorem 2.} {\it Fixed $\O,s>0$ let
$C(\h)=\O e^{-s\hdm}$; and let $\NN=\g\hdm$, $\g>0$. For $\h$ small and
$|\e|<\e_0$, where $\e_0$ is given by \equ(4.3), the Hamiltonian
\equ(1.1) has, for all $\h$ small enough, a family of hyperbolic
invariant tori, with stable and unstable manifolds of dimension $3$,
whose rotation vectors $\oo'=(\hdp\o_1',\hdm\o_2)\in\SX_0$
verify \equ(3.5) and therefore have first components which fill
the interval $[\tilde\o_a,\tilde\o_b]$ within
$O(\O e^{-s\hdm})$ for all $\h$ small enough, provided $\e<\e_0$.

More precisely the parametric equations of such tori can be written,
for $\e$ small enough and
$\oo'\in \CS_0$ verifying \equ(3.5), as:

$$\eqalign{\AA=&\V H_{\oo'}(\aa')\cr\aa=&\aa'+\V h_{\oo'}(\aa')\cr}
\qquad
\eqalign{I=&\II_{\oo'}(\aa')\cr \f=&\Psi_{\oo'}(\aa')\cr}\Eq(5.4)$$
with $\V H_{\oo'},\V h_{\oo'},\II_{\oo'},\Psi_{\oo'}$,
at fixed $\oo'$, analytic in $\e$ and in $\aa'\in{\bf T}^2$
and divisible by $\e$, so that such functions can be found by
perturbation expansions.  The smallness condition on $\e$ is $|\e|<
O(\e_0 [C(\h)]^{q\g^{-1}\hdp})$ (a very weak condition, as $\g$ can be
arbitrarily prefixed) and the motion of the data \equ(5.4) is simply
$\aa'\to\aa'+\oo' t$.}
\*

It is interesting to write the complete
condition that $\e$ is small enough (including the constants) and
a more quantitative expression for the filling of the action axis.
Since the above theorem relies on [CG] one has to find bounds for
several quantities associated with \equ(4.4). If $M_0^{-1}$ denotes
a lower estimate for the minimum value of the second derivative of
\equ(1.1) evaluated at $\e=0$
with respect to $a=\hdp A_1$ (\ie if $M_0$ is defined as in
\equ(2.2)) one finds, applying the condition (5.90) of [CG], for some
$B,q>0$:

$$|\e|<\fra1B \e_0 \,\Big(\fra{\l_0\G_0}{E_0}\, \fra{\l_0^2\h}{E_0
M_0}\,\x_0\,C(\h)\Big)^{q/\NN}= \hbox{const.}\,\e_0\,
[C(\h)]^{q\g^{-1}\hdp}=O(\h^{\fra92+}) \Eq(5.5)$$
having unified various constants to simplify the expression (and after
translation to the present symbols and conventions) and denoting
$\fra92+$ a prefixed number bigger than $\fra92$.
And under the same conditions the filling
of phase space is within $O(M_0E_0\l_0^{-1} C(\h))$.

A more detailed description of the above analysis is the
following strengthening of the above theorem.

\*

\0{\cs Theorem 3.} {\it Under the same hypotheses of theorem 2 above
there exists a $C^\io$ function $\lis\HH_0(\AA',x')$, $x'=p'q'$, defined
in $\DD_{\r_0/2,\x_0/2,\k_0/2}$ and a $C^\io$ canonical
transformation, defined on the latter domain, and having the form:

$$\eqalign{
\AA=&\AA'+\V H(\AA',\aa',p',q'), \qquad I=\II(\AA',\aa',p',q')\cr
\aa=&\aa'+\V h(\AA',\aa',p',q'), \qquad
\f=\Psi(\AA',\aa',p',q')\cr}\Eq(5.6)$$
with $\V H,\V h,\II,\Psi$ of class $C^\io$ divisible by $\e$ and
analytic in $\e,\aa'$ at fixed $\AA',x'=$ $p'q'$ such that if
$\oo'=\dpr_{\AA'} \lis\HH_0(\AA',x')$ verifies \equ(3.5) then in the new
coordinates the Hamiltonian coincides, together with its first order
derivatives, with $\lis\HH_0(\AA',x')$. And $\lis\HH_0$ is close within
$O(\e)$ to $h(\hdp A'_1)+\hdm \o_2 A'_2+G(p' q',\hdp A'_1)$.  Hence the
motion of the data $(\AA',\aa',p',q')$ for which $\oo'$ verifies
\equ(3.5) are $\AA'\to\AA'$, $\aa'\to\aa'+\oo' t, \, p'\to p' e^{-g'
t},\, q'\to q' e^{+g' t}$ with $g'=\dpr_{x'}\lis\HH_0(\AA',x')$ if
$x'=p'q'$.}

\*

\0{\it Remarks.}
(1) The value $C(\h)= \O e^{-s \hdm}$ is particularly interesting
in view of the results in [GGM1], where the Hamiltonian
$$ \HH = \hdp A_1  + \h \fra{A_1^2}{2} + \hdm A_2
+ \fra{I^2}{2} + g^2(\cos\f-1)+\e f(\f,\a_1,\a_2) \Eq(5.7) $$
is studied, with $f$ an even trigonometric polynomial of degree
$N,N_0$ in $\aa,\f$ respectively, and the
homoclinic splitting of the invariant tori whiskers is found to be of
size of order $C(\h)|_{s=\p/2g}$.  Note that \equ(5.7) with $(I,\f)$
near $(0,0)$ is a particular case of \equ(1.1). One just uses
``Jacobi's coordinates'' $(p,q)$ instead of $(I,\f)$, see
Appendix A9 of [CG].

In [GGM1] the slow frequency is considered to be $\h^a$, with $a\ge
0$, so that $a=1/2$ is only a particular case.  Of course in what
follows we could also consider an arbitrary value $a\ge0$, and
essentially nothing would change, but we have preferred to confine
ourselves to the case $a=1/2$ for definiteness.

(2) In general the functions $\V H,\V h$ in theorem 3
do not have zero average for $p'=q'=0$.
The even symmetry of $f$ and of the pendulum Hamiltonian in \equ(5.7)
imply that the variables $\AA'$ {\it have a simple physical
interpretation} for the points on the invariant tori: they are just
the {\it time averages} of the unperturbed actions.  In fact this
``{\it absence of torsion}'' is a very remarkable and useful property
of the system \equ(5.7) ({\it Thirring model}) which was pointed out
in [G3,G4] (and called the property of the tori of being {\it
twistless}). It is due to the special symmetries of the system
\equ(5.7) and to the separation of the energy into a quadratic part
involving actions only and an angular part involving only the angles.

(3) A further consequence of the symmetries of \equ(5.7) is that each
invariant torus described by theorem 2 has stable and unstable
manifolds that intersect along a trajectory, a {\it homoclinic
trajectory}, that when $\f=\p$ has angular coordinates $\aa=\V0$.

The difference at $\aa,\f$, that we denote $\V Q(\aa,\f;\hdp A'_1)$,
between the $\AA$--coordinates of the stable and unstable manifolds of
the torus $\TT(A'_1)$ with average action $ A'_1$, vanishes at
$\f=\p,\aa=\V0$, see [CG], \S9. Hence we say that $\aa=\V0,\f=\p$ is a
{\it homoclinic point} for $\TT(A'_1)$. Setting $\V Q(\aa)\defi\V
Q(\aa,\p;\hdp A'_1)$ the matrix $D_{ij}=\dpr_{\a_i}
Q_j(\aa)|_{\aa=\V0}$ is called the {\it intersection matrix} and in
general has a determinant $\det D$ that does not vanish:
its value is called the {\it homoclinic splitting}, see [GGM1].

(4) The same result as theorem 2 can be obtained, for the Hamiltonian
\equ(5.7), also from the Lindstedt's series expansion for the
stable and unstable manifolds of the invariant tori, by using the
methods developed in [E,G3,Ge1,Ge2] and an (improved) version of
Siegel--Bryuno's lemma; but this is a completely different approach to
the problem and we expose it, in detail, in a forthcoming paper, [GGM2].
The new approach yields, as expected, a better bound on the convergence
radius $\e_0$: namely $\e_0=O(\h^2)$ compared to the $\sim O(\h^{4.5+})$
that would be obtained by applying theorem 1 to \equ(5.7) (see the
remark 1 after theorem 1 and the comments after \equ(5.5)).

(5) Theorem 3 means that there is a canonical system of coordinates in
which the Hamiltonian takes the {\it normal form} $\lis\HH_0$ given by
an expression like \equ(3.3) {\it on a set with very small complement,
in general not empty and open} (we stress that this is a statement
different from the one that would hold if the Hamiltonian could have
been really put in the form \equ(3.3), \ie in the whole vicinity of
the real domain of \equ(1.1)).  The reason $\lis\HH_0$ is called a
``normal form'' is that the data with $\oo'$ verifying \equ(3.5)
evolve in a very simple fashion \ie as $\AA'={\rm const.}$,
$\aa'\to\aa'+\oo' t$, $p'\to p' e^{-\lis g 't}$ and $q'\to q'e^{\lis g
't}$ (see \equ(3.3) and
\equ(3.4)).

(6) Another consequence of theorem 3 is that fixed $\AA'$ so that
$\oo'=\dpr_{\AA'} \lis\HH_0(\AA',0)$ verifies \equ(3.5) it is possible to
fix $A_1(x')$ so that $\dpr_{\AA'} \lis\HH_0(\AA'(x'),x')\= \oo'$ for,
say, $|x'|<\k_0/4$.  And, at the same time, the energy of the motions
that start in $\DD_{\r_0/2,\x_0/2,\k_0/2}$ with $\AA'=\AA'(x')$ and
$p',q'$ with $p'q'=x'$ and with $\aa'\in{\bf T}^2$ arbitrary is
$x'$-independent.
Here the linearity in $A'_2$ is used to adjust the energy
once $A'_1(x')$ is determined: so that the apparently useless role of
the variable $A'_2$, ``reservoir energy'' for keeping the second
rotator in a constant speed rotation, can be well appreciated here.
This is a key property for the theory of diffusion in phase space
developed in [CG].

\*

\0{\bf\S6. Existence of heteroclinic chains.}
\numsec=6\numfor=1\*

The novelty of the results in \S 5 is the possibility of using the
``large angles theorem'' of [GGM1] in the same way as the erroneous
result (with the same name) in \S10 of [CG] was used to discuss
heteroclinic chains.

In this section we restrict considerations to the system described by
the Hamiltonian \equ(5.7).  Given an invariant torus with average
action $\AA'$ such that the corresponding
rotation vector $\oo$ verifies \equ(5.2) with $C(\h)=\O e^{-s\hdm}$,
$\O,s>0$, one has that the torus and its stable and unstable manifolds
(whiskers) are analytic for $|\e|<2\e_0$, for a suitable $\e_0$.  The
discussion in \S5 gives $\e_0=O(\h^{\fra92+})$ (by
taking $\NN=\g\hdm$ with $\g$ large enough, see \equ(5.5)),
by the argument leading to theorem 2; as mentioned in \S5, remark 4, a
better value $\e_0=O(\h^2)$ can be obtained with a different method.

Then we want to infer that the homoclinic splitting, at the homoclinic
point $(\f,\a_1,\a_2)=(\p,0,0)$ (see \S5, remark 3), of such a torus
is generically given by the first order ({\it Melnikov's integral}),
\ie it is of the form:

$$\s\e^2\h^{-b}e^{-\fra\p2 |\o_2|g^{-1}\hdm}\Eq(6.1)$$
with $\s$ a non vanishing constant and $b$ a positive constant
linearly depending on the degree $N_0$
(in $\f$) of the perturbation $f$. The
finiteness of $N_0$ is thus a key assumption for the validity of the
following analysis.

The theorems of [GGM1] do not apply directly because in the latter
paper $C(\h)$ was taken proportional to $\h^d$ for some $d>0$.  On the
other hand, by the considerations in \S5, the results of [GGM1] can be
immediately extended to cover also this case, \ie to conclude that the
splitting is given asymptotically by Melnikov's integral,
as we shall see below.

By theorems 1$\div$3 above, the invariant tori and their
stable and unstable manifolds are analytic in $\e$ for $|\e|<2\e_0$,
with $\e_0=O(\h^{c'})$, where $c'>9/2$ or $c'=2$ (according to which
bound for $\e_0$ is chosen, see the remark 4 in \S5).
This implies that, to any order $h$, the parametric equations of the
manifolds (\ie $\hh,$ $\V H$ as functions of $\aa,\e$ at $x'=p'q'=0$,
$A'_1$ fixed) evaluated at $\f=\p$ are analytic in $\e$ and their
Taylor series coefficients can be bounded to order $h$ by:
$$ D_2 B_2^{h-1} (\e 2^{-1}\e_0^{-1})^h \Eq(6.2) $$
for some positive constants $D_2,B_2$.

Moreover we know, from the analysis in [G3] and [GGM1], that, to any
order $h\le\NN=\g\hdm$, the Taylor coefficients for the splitting are
bounded, to order $h$, by
$$ D_3 h!^4 B_3^{h-1}(\e\h^{-\b'})^{h} e^{-\fra\p2|\o_2|g^{-1}\hdm}
\Eq(6.3) $$
with $\b'=2(N_0+1)+2$, for some positive constants $D_3,B_3$ and with
$N_0$ equal to the degree in $\f$ of the perturbation $f$ (hence
$N_0\le N$: the convenience in distinguishing between $N$ and $N_0$ is
that it is possible to extend the present work to cover cases in
which $f$ is analytic in $\aa$, \ie $N=\io$, provided one keeps it
polynomial in the angles $\f$).

{\it No factor $C(\h)$ appears in \equ(6.3) as no resonances occur in
the bounds of the divisors that appear in the perturbation expansions}
(of [G3,GGM1]) contributions to the splitting at orders $\le \NN$, by
the very choice of $\NN$. In fact as long as only orders $\le \NN$ are
considered one has $|\oo'\cdot\nn|\ge \hdp$ as $|\nn|\le N\g\hdm=N\NN$.

Then if $\e\le O(\h^{c})$, $c>\max\{c',\b'+2\}$ $=$ $\b'+2$, the
contribution to the splitting of the orders $>\NN$ can be estimated
from \equ(6.2) by $O((\e 2^{-1}\e_0^{-1})^\NN)$: which is much smaller
than the generic homoclinic splitting, \equ(6.1), evaluated from
Melnikov's integral.

The contribution arising from the orders $3\le h\le\NN$ is bounded, via
\equ(6.3), by
$O((\e\h^{-\b})^{3}e^{-\fra\p2|\o_2|g^{-1}\hdm})$, with $\b=\b'+2$,
see [G3,GGM1], also much smaller than \equ(6.1) if $\e$ is smaller
than a high enough power of $\h$ (\ie {\it without a condition of
exponential smallness of} $\e$).

Hence we conclude that under a condition like $|\e|<O(\h^{c})$, with
$\h$ small enough, tori with average action $\AA'$, arbitrarily
prefixed energy {\it and} verifying \equ(5.2) with $C(\h)=\O
e^{-s\hdm}$, $\O,s$ preassigned) exist and their homoclinic splitting
has size $O(\e^2\h^{-b}e^{-\fra\p2|\o_2|g^{-1}\hdm})$ generically in
perturbations $f$ which are trigonometric polynomials in $\f$ (and the
constant $b$ depends on the degree $N_0$ in $\f$ of $f$).

For instance the simplest case \equ(5.7) with
$f=\cos(\a_1+\f)+\cos(\a_2+\f)$ and $g=1$ gives a leading splitting,
see [GGM1] eq. (6.14) and (7.19), as $\h\to0$:

$$32 \p\hdm \e^2 e^{-\fra\p2\hdm}\Eq(6.4)$$

We also conclude the (generic) existence, for $\e<O(\h^{c})$ and $\h$
small enough, of chains of isoenergetic invariant tori
(``{\it heteroclinic chains}'') with average actions $\AA_j'$ such that:

\0(i) $\oo'(0,\hdm A_{j1}')$ verify \equ(5.2) with
$C(\h)=\O e^{- s\hdm}$ with $s,\O$ that can be fixed {\it a priori}, and
$|\AA'_j-\AA'_{j+1}|<O(\h^{-c''}e^{- s\hdm})$ for a suitable $c''$;

\0(ii) the homoclinic splitting has size
$O(\e^2\h^{-b}e^{-\fra\p2|\o_2|g^{-1}\hdm})$ (\ie far larger than
$C(\h)$ if $s$ is fixed \ap to be $s>\fra\p2 g^{-1}$);

\0(iii) the unstable manifold of the torus
with average action $\AA'_j$ intersects the stable manifold of the
torus with average $\AA'_{j+1}$ in a {\it heteroclinic trajectory}
with splitting at $\f=\p$ of order
$O(\e^2\h^{-\b}e^{-\fra\p2|\o_2|g^{-1}\hdm})$: this follows, as usual,
from an application of the implicit functions theorem
because the equation to be solved for finding a heteroclinic point is
an implicit equation with Jacobian determinant, at the trivial
solution $\aa=\V0$ (corresponding to the homoclinic point at $\f=\p$
of $\AA'_j$), given by the splitting $\det D$, see also [CG,GGM1];

\0(iv) the ``genericity'' is a very explicit condition because it simply
requires the lowest order value of the splitting to be non vanishing:
this holds generically in $f$ (picked up inside the class of functions
we are considering);

\0(v) furthermore, since the set of points verifying \equ(5.2)
is abundant in the sense of the above theorems 2 and 3, we also obtain
(at no extra cost, \ie by ``abstract reasoning'' on measure theory)
that the values of $A_{j1}'$ are density points for the set of values
$A'_{j1}\in[-\hdm R,\hdm R]$ whose corresponding rotation
vectors verify \equ(5.2) with the chosen value
of $C(\h)$; and we can suppose that the values of $A'_{j1}$ at the
extremes of the chain are close to the extremes of the interval
$[-\hdm R,\hdm R]$ within $\e$ (or even closer).

We can summarize the above discussion in the following result.

\*

\0{\cs Theorem 4.} {\it Given the system described by the
Hamiltonian \equ(5.7), there is a sequence of invariant tori
with average actions $\AA_1',\ldots,\AA_\NN'$, with
$|\AA_1'-\AA_\NN'|\simeq 2R\hdm$ (within $O(\e)$) and
$|\AA_i'-\AA_{i+1}'|\le Re^{-s\hdm}$ for $i=1,\ldots,\NN-1$ and $s>0$,
such that the unstable manifold of each invariant torus intersects the
stable manifold of the torus following it along the chain. At the
heteroclinic point at $\f=\p$ the splitting between the two manifolds
is given by $\s'\,\e^2\h^{-b}e^{-\fra\p2|\o_2|g^{-1}\hdm}$, where
$\s',b$ are constants depending on the perturbation $f$, with $\s'$
generically not vanishing.}

\*
This extends the theorem of existence of heteroclinic chains
discussed in [GGM1], \S 8, to the anisochronous case.

\*
\0{\bf\S7. Fast averaging theorem and abundance.}
\numsec=7\numfor=1\*

A second remarkable application is a fast averaging theorem which
follows from the proof of theorem 1 and concerns a system with
Hamiltonian:
$$\HH=h(\hdp A_1)+\hdm\o_2A_2+\fra{I^2}{2J_0}+J_0 g^2(\cos\f-1)+
\e_1 f_1(\aa,\f)+\e_2f_2(\aa,\f)\Eq(7.1)$$
where $I,\f$ are conjugate variables, $J_0,g$ are strictly positive
constants and $\e_1,\e_2$ are perturbation parameters. The function
$f_1$ is {\it monochromatic} (\ie with Fourier transform not vanishing
only on modes different from zero and multiples of a fixed
$\nn_0=(\n_1,\n_2)$ with $\n_2\ne0$), while $f_2$ is only required to
be a trigonometric polynomial.

On the basis of the results of the previous sections we can expect that
in order to have ``long'' heteroclinic chains of invariant tori we need
to require $\e_1$ and $\e_2$ to be {\it both} small (of the order of
$\e_0$ of theorems 1$\div$4).

{\it But this is not the case} because of the monochromatic nature of
$f_1$ and we shall show that the results of theorem 1$\div$3 hold
essentially under the assumptions that $\e_2$ is small of order $\e_0$
while $\e_1$ can be quite large, up to almost $\hdm$: in particular
the value $\e_1=1$ is amply allowed. In a sense also theorem 4 could
hold; however we can only show (see below) that verification of its
validity can be reduced, at least in principle, to a finite
computation if $\h,\e_2$ and $\e_1=1$ are fixed with $\e_2<\e_0$.

{\it We consider first the case $\e_2=0$.}  In this case, by performing
the canonical transformation with generating function
$S=\AA'\cdot\aa+I'\f-\e_1\hdp \o_2^{-1}\dpr_2^{-1} f_0(\aa,\f)$, one sees
that the size of the perturbation can be reduced to $O(\hdp\e_1)$ under
the condition that $\hdp\e_1$ is small enough (so that the canonical map
generated by $S$ makes sense in a domain slightly smaller than the
domain of $h$).

This is {\it remarkable} because the change of coordinates is {\it
globally defined} and it is not restricted to the vicinity of $I=\f=0$,
and it leaves the perturbation still a monochromatic trigonometric
polynomial with zero average over $\aa$. This cannot be pursued to
higher orders without losing the trigonometric polynomial
nature of the perturbation.

Then we can then use Jacobi's coordinates near $I=\f=0$, see [G2], to put
the Hamiltonian in the form \equ(1.1) with $\e f$ of order $O(\hdp\e_1)$.

It follows from the proof of theorem 1 (see the remark 2 in \S4) that
if the perturbation is monochromatic the condition for being able to
cast the Hamiltonian in the form \equ(4.4), with a prefixed and $\h$
independent $\NN$, say $\NN_0$, is that $|\e_1|<O(1)$, so that we see
that \equ(7.1) can be cast in the form \equ(4.4) with a prefixed
$\NN_0$ for $|\e_1|\hdp$ small enough.

We consider now the case $\e_2\ne0$ and we imagine performing the
change of variables that puts the part of \equ(7.1) {\it not
containing the last term} into the form \equ(4.4) for some $\NN_0$.

This casts the full Hamiltonian \equ(7.1) into the form \equ(1.1) {\it
except for the fact that the perturbation is no longer} a trigonometric
polynomial in the $\aa$ variables: it is just an analytic function in
a domain of size $\x_0>0$ in the complex planes for the $\aa$
variables.

We truncate its Fourier expansion in the $\aa$ at $|\nn|<N$, where $N$
is left as a parameter. The remainder can be bounded in its domain by
$a e^{-\x_0 N}$ for some $a>0$ of order $O(|\e_1\hdp|+|\e_2|)$.

We are, therefore, in a position to apply again theorem 1 and to cast
the full \equ(7.1) into the form \equ(4.4). We make the following
choices: $\NN=\hdm \sqrt{(\log\h^{-1})^{-1}}$, $N=\fra12\hdm
\sqrt{\log\h^{-1}}$, (assuming $\h<\fra12$, which is not restrictive
as we are interested in properties holding as $\h\to0$).

Combined again with theorem 1 above, this means that we can put the
Hamiltonian in the form \equ(4.4) with the last term replaced by
an ``effective interaction'' of order:
$$\Big(\fra{((\e_1\hdp)^{\NN_0}+{\e_2})(\log\h^{-1})^{-\fra32}}
{\h^{9/2}\lis\e_0}
\Big)^{\hdm \,(\log\h^{-1})^{-1/2}}+ a e^{-\fra{\x_0}2\hdm
(\log\h^{-1})^{1/2}}\,\ll\, e^{-z\hdm}\Eq(7.2)$$
provided $\e_2=O(\h^c)$, with $c$ large enough and $\NN_0$ so prefixed
that $\NN_0/2\ge c$; $a,\lis\e_0$ are constants that (if wished) can
be explicitly estimated from the above argument.  The inequality holds
for {\it any} prefixed $z>0$ provided $\h$ is correspondingly suitably
small.

This means that, in some sense, the size of the monochromatic term in
\equ(7.1) does not matter too much and $\e_1=1$ (or even almost
$\e_1\simeq \hdm$) is a sufficient condition, together with
$|\e_2|<\e_0$ and $\h$ small enough, to guarantee the existence of an
invariant torus which is run quasi-periodically with a rotation $\oo'$
verifying a (very weak) Diophantine condition
($C(\h)=\O e^{-s\hdm}$) and whose equations can be constructed by a
convergent perturbation series.

A further consequence is that the Hamiltonian \equ(7.1) has many
invariant tori whose average actions fill the action space as
described in theorem 2 above, so that we can apply the results of
[GGM1].  We summarize the above discussion in the following
averaging theorem.

\*

\0{\cs Theorem 5.} {\it Given $s>0$ consider the
Hamiltonian \equ(7.1) with $f_j$ verifying the assumptions following
\equ(7.1). If $|\e_1|<\h^{-d}, \,|\e_2|<\h^{+d'}$ with prefixed
$0\le d<\fra12$ and $d'>\fra92$ there is a family of invariant tori
$\TT(\AA')$ with average actions $\AA'$ filling action space
within $O(e^{-s\hdm})$. The parametric equations of such tori can be
computed by perturbation theory (\ie by convergent power series in
$\e_1,\e_2$) and they are, together with the homoclinic splitting,
analytic functions of $\e_1,\e_2$ provided $\h$ is small enough
(depending on the size of $s,d,d'$).}
\*

Note that the ``torsion free'' property allows us to fix the rotation
vector of the motions on an invariant torus by fixing the average
value $A'_1$ of the action, no matter which is the perturbation size
$\e$: in fact in the model \equ(7.1) that we are considering it is
$\oo'=(\hdp+\h A'_1,\hdm\o_2)$ which is $\e$--independent (remarkably
enough).

Apart from the intrinsic interest of theorem 3 it has some relevance
for the theory of drift and diffusion in {\it a priori} stable
systems, like the precession problem whose analysis was attempted
in [CG], \S12. We discuss this aspect in the next section.
\*

\0{\bf\S8. Drift and diffusion under large monochromatic forcing
and small quasi periodic perturbation, or in a priori stable systems}
\numsec=8\numfor=1\*

The interest of the above theorem 5 is, again, that it {\it might} be
used, by repeating the argument in \S6 to show the existence of
heteroclinic chains with the same properties of the ones discussed in
that comment even if $\e_1=O(1)$.

This time {\it there is, however, an extra difficulty if $\e_1$ is
large, \ie really of order $1$}: the argument of \S6 provides us
with a tool to bound the remainders, but {\it not} to find out which
is the dominant term among those of ``small'' order (\ie smaller than
$\NN$ in \S6).

The reason is simply that the contributions of higher order in $\e_1$
not only are not smaller than those of order $1$ but in fact are
checked to be \ap larger. Meaning that one can find {\it some}
contributions to them coming from suitable graphs, in the sense of the
representation in [GGM1], and {\it larger} than the lowest order
contribution: but in a subject where cancellations are quite common
this does not seem to mean much.

{\it However the results in [GGM1] and theorem 4 above show that the
problem of proving the existence of a heteroclinic chain joining the
extremes of the action $A'_1$ interval $[-R\hdm,R\hdm]$ can in
principle be solved by a finite calculation}, if $\e_1$, $\e_2$
and $\h$ are fixed. The program is illustrated in the following.

In fact such chains can be proved to exist and to consist of finitely
many elements for $|\e_1|$ small enough (by theorem 3) and the above
discussion shows that the series in $\e_1,\e_2$ for the splitting is
analytic, and the term of order $h_1$ in $\e_1$ and $h_2$ in $\e_2$ is
bounded by
$$ B^{h_1+h_2} (\e_1\h^{1\over2})^{h_1}(\e_2\h^{-c})^{h_2} \Eq(8.1) $$
where $B$ is a suitable constant.

Assume that we can show that the sum of the first $h_1+h_2\le
\g\hdm$ terms in the series for the splitting is bounded
from above and below by a quantity $O(e^{-\fra\p2|\o_2| g^{-1}\hdm})$;
then the remainder is surely smaller than this quantity.  Hence we can
approximate the heteroclinic points for an allowed value of $\e_1$:
keeping fixed the average action $A'_1$ we can increase it following
the evolution of the heteroclinic point. This is a finite calculation,
in principle, because we start from $\e_1>0$ and have convergent
expansions up to $\e_1=1$ (and beyond up to almost $\hdm$) thanks to
the above fast averaging observations, valid close enough to the tori
(where the changes of variables discussed to reach \equ(7.2) make
sense). Thus with a finite calculation one can get controlled
approximations of the values of the heteroclinic angles (at fixed
$\e_1,\e_2,\h$).

What is not \ap guaranteed is that such angles, {\it which start
positive} and ``large'' compared to the separation between the tori,
do not become small while $\e_1$ grows.

However there is no reason for this to happen: one expects the angles
to become wider, not smaller, or to vanish only for finitely many
values of $\e_1$, at least generically.

Hence a computer assisted analysis of the problem is likely to succeed
and it might even be used to reach values of $\h$ and $\e_1,\e_2$
where the angles are large and the spacing between the tori can be
widened (thus diminishing the number of invariant tori of the
chains). As discussed in [CG] the drift along a heteroclinic chain
with non zero splitting is always present and one would obtain models
with Arnol'd diffusion in systems like
\equ(7.1) with $\e_1=1$. This would correct even the applications of
the incorrect version of the large angle theorem of [CG] (see \S12 of
[CG]) and make it rest entirely on the corrected version in [GGM1].

In fact a problem like \equ(7.1) with $\e_1$ large ($\e_1=O(1)$)
arises in the theory of Arnol'd diffusion in \ap stable systems, like
the precession problem treated in \S12 of [CG]. From the analysis in
[CG] it seems that such large (but monochromatic) coupling Hamiltonian
systems might arise often in the reduction of a motion near a
multiple resonance of a \ap stable system.
\*
\0{\it Remark.
It is interesting to note that the above program with $h(a)=\o_1 a$,
\ie \equ(7.1) in the isochronous case, can be carried out without need
of a computer as remarked in [GGM1] (the diffusion then follows via
the results in [G2]) at least for all $\e_1<O(\hdm)$ except possibly a
finite number of them.}
\*

\0{\bf Acknowledgments.}
{\it This work is part of the research program of the European Network
\# ERBCHRXCT940460 on: ``Stability and Universality in Classical
Mechanics",}
\*
\ifnum\mgnf=0\pagina\fi
\0{\bf Appendix A1. Proof of theorem 1.}
\numsec=1\numfor=1\*

Suppose that for $h\le k-1$ one has:
$$||\F^{(h)}||\le D B^{h-1} (h-1)!,
\qquad ||\tilde G^{(h)}||\le E_0 B^{h-1} (h-1)!
\Eqa(A1.1)$$
where $D$ is such that the bound holds for $h=1$. Here $||\cdot||$
denotes the maximum in $\DD_{\r_h,\x_h,\k_h}$ with
$\x_h=\x_0-h\d,\r_h=\r_0 e^{-\d h},\k_h=\k_0e^{-h\d}$.  Furthermore
the Fourier transform of $\F^{(h)}$ with respect to the
$\aa$--variables has only modes $\nn$ with $|\nn|<N h$.

Therefore for $h<\z \h^{-1}N^{-1}$,
with $\z=|\o_2|(2\tilde\o_b)^{-1}$,
we can bound the various terms in
the difference between the two sides of \equ(3.1) as follows.

We expand in Taylor's series the functions in \equ(3.1) with respect to
the point obtained by setting $\F=0$. We define $\oo\defi(\hdp \dpr
h(\hdp A'_1)+\hdp \dpr G(p'q,\hdp A'_1),\hdm\o_2)$, $\lis g(p'q,\hdp
A'_1)\defi\dpr_x G(x,\hdp A'_1)|_{x=p'q}$, $\ndpr\defi q\dpr_q-
p'\dpr_{p'}$ and $\D\defi \oo\cdot\dpr_\aa+\lis g\ndpr$. Then the
{\it linear term} in $\F^{(k)}$ is simply $\D \F^{(k)}$.

The nonlinear terms arising from $h(\hdp(A'_1+\dpr_{\a_1}\F))$
contribute to $\F^{(k)}$ an amount $X_1^{(k)}$ such that:

$$\D X_1^{(k)}=\Big(\sum_{a\ge2}
h^{(a)}\hdp^a\sum_{k_1+\ldots+k_a=k\atop
k_j\ge1}\prod_{j=1}^a\dpr_{\a_1}\F^{(k_j)}\Big)^*\Eqa(A1.2)$$
where the $*$ means that from the Taylor expansion in $p',q$ and
Fourier expansion in $\aa$ of the term in parentheses one has to
subtract the terms which have $\V0$ Fourier mode and that have an
equal power of $p'$ and $q$. In fact the latter terms make up the
contribution to $\tilde G^{(k)}$ of the nonlinear terms arising from
$h(\hdp(A'_1+\dpr_{\a_1}\F))$. After this subtraction the operator
$\D$ is invertible, \ie $X_1^{(k)}$ is well defined and the action of
$\D^{-1}$ can be bounded by $\G_0^{-1}C(\h,\d)$ because as long as
$|\nn|<\z\h^{-1}$ one has:
$$\sum_{\nn,n,m}^+ \fra{e^{-\fra\d2( |\n_1|+|\n_2|+|m|+|n|
)}}{(|\n_1|\hdp+|\n_2|\hdm+|m-n|)\G_0}\le \G_0^{-1} C(\h,\d)
\Eqa(A1.3)$$
where the $+$ means that the sum runs over the $\nn,m,n$ for which the
divisor does not vanish and, furthermore, $|\nn|<\z\h^{-1}$,
so that no ``resonances'' occur. The quantity $C(\h,\d)$ can be taken
$$C(\h,\d)=c\left(\hdp \d^{-3}+\hdm
\d^{-2}+\d^{-1}\right)\log \d^{-1}>1 \Eqa(A1.4)$$
where $c$ is a suitable constant (recalling also that $\e,\h<1$, see \S2).

Making use of the analyticity assumption, which implies that
$|h^{(a)}|< E_0\r_0^{-a}$ in $D_{\r_0,\x_0,\k_0}$, one has
in $D_{\r_k,\x_k,\k_k}$:
$$|X_1^{(k)}|\le \fra{E_0 C(\h,\d)}{\G_0}\sum_{a\ge2}\Big(\fra{2\hdp
D}{\r_0\d}\Big)^a B^{k-a}
\sum_{k_1+\ldots+k_a=k\atop k_j\ge1}\prod_{j=1}^a (k_j-1)!\Eqa(A1.5)$$
where the factor $2$ above comes from the estimates
of the Fourier transform of the factors $\dpr_{\a_1}\F^{(k_j)}$
in \equ(A1.2) inside the domains
$D_{\r_{k_j+1/2},\x_{k_j+1/2},\k_{k_j+1/2}}$.

We shall repeatedly use below the inequality:
$\sum_{k_1+\ldots+k_a=k\atop k_j\ge1}\prod_{j=1}^a (k_j-1)!\le (k-1)!$
(to prove it, simply bound the product of the factorials by
the $(k-a)!$, and the number of addends by $(k-a+1)^{a-1}/a!$).
Then, from \equ(A1.5),
$$|X_1^{(k)}|\le \fra{E_0 C(\h,\d)}{\G_0}
B^{k-1}(k-1)!\, B\,\big(\fra{2\hdp D}{\r_0 B\d}\Big)^2 2
\qquad {\rm if } \quad \fra{2\hdp
D}{\r_0 B\d}<\fra12 \Eqa(A1.6) $$
Hence in $D_{\r_k,\x_k,\k_k}$:
$$|X_1^{(k)}|<\fra15 D B^{k-1} (k-1)!\qquad {\rm if} \quad
\fra{2\hdp D}{\r_0 B\d}<\fra12,\qquad 2\fra{E_0 C(\h,\d)}{\G_0}
\fra{4\h D}{\r_0^2 B \d^2}<\fra15\Eqa(A1.7)$$
The higher order terms contribution $X_2^{(k)}$ from the third term
in \equ(3.1) is:
$$\D X_2^{(k)}=
\Big(\sum_{n,a\atop a+n\ge2} G^{(a,n)} \sum_{k_1+\ldots+k_a+\atop
h_1+\ldots+h_n=k} {\h^{\fra12}}^n\prod_{j=1}^n\dpr_{\a_1}\F^{(h_j)}
\prod_{i=1}^a q\dpr_q\F^{(k_i)}\Big)^*\Eqa(A1.8)$$
with $h_j,k_i\ge1$. So that in $D_{\r_k,\x_k,\k_k}$:
$$\eqalign{
&|X_2^{(k)}|\le \fra{E_0 C(\h,\d)}{\G_0} \sum_{n,a\atop a+n\ge2}
\fra{2^n\h^{\fra12 n}}{\r_0^n\d^n}D^n \k_0^{-a} B^{k-n-a}
\fra{2^aD^a}{\d^a}\,
(k-1)!\le\cr &\le \fra{E_0 C(\h,\d)}{\G_0}
\sum_{n,a\atop a+n\ge2}\Big(\fra{2D}{\l_0 B\d}\Big)^{n+a}
{\h^{\fra12 n}}B^k (k-1)! \le \fra15 D B^{k-1} (k-1)! \cr} \Eqa(A1.9)$$
provided:
$${2D\over \d\l_0 B}<{1\over 2},\qquad 4\fra{E_0 C(\h,\d)}{\G_0}
\fra{4D}{\l_0^2 B \d^2}<\fra15\Eqa(A1.10) $$
and note that \equ(A1.10) implies the conditions in \equ(A1.7).

Likewise the fourth term in \equ(3.1) yields a higher order contribution
$X_3^{(k)}$ bounded as:
$$\eqalign{&
|X_3^{(k)}|\le \fra{E_0 C(\h,\d)}{\G_0}\sum_{a,n\atop a+n\ge1}
\fra{\hdp^a}{\r_0^a\k_0^{\fra12 n}}
\fra{2^{a+n}D^{a+n} B^{k-1-a-n}}{\d^{a+n}\k_0^{\fra{n}2}} (k-1)!\le
\cr
&\le 4 B^{k-1} (k-1)!\fra{E_0 C(\h,\d)}{\G_0} \, {2 D\over \l_0 B\d}
\qquad {\rm if} \quad \fra{2 D}{\l_0 B \d}<\fra12\cr}\Eqa(A1.11)$$
so that one has in $D_{\r_k,\x_k,\k_k}$:
$$ |X_3^{(k)}|\le \fra15 D B^{k-1}(k-1)!, \qquad {\rm if}\quad
\fra{2D}{\l_0 B\d}<\fra12,\qquad 4\fra{E_0 C(\h,\d)}{\G_0}
{2\over\l_0B\d} <\fra15\Eqa(A1.12)$$

The fourth higher order contribution comes from the third term in the
r.h.s. of \equ(3.1) and is bounded in the same way as $X_2^{(k)}$,
leading to the same bound under the same conditions.
The fifth higher contribution
comes from the last term in the r.h.s. of \equ(3.1), which, up to
the term $\tilde G^{(k)}$, is bounded in $D_{\r_k,\x_k,\k_k}$ as:
$$\eqalign{
&|X_5^{(k)}|\le \fra{E_0 C(\h,\d)}{\G_0}\sum_{n\ge 1}
\k_0^{-n} B^{k-n} 2^n D^{n}\fra1{\d^n} (k-1)! \le\cr &\le
2 \fra{E_0 C(\h,\d)}{\G_0}\fra{2D}{\l_0B\d}
B^{k-1}(k-1)!\le \fra15 D B^{k-1}
(k-1)!\cr}\Eqa(A1.13)$$
provided:
$$\fra{2D}{\l_0B\d}<\fra12 \qquad 2 \fra{E_0 C(\h,\d)}{\G_0}
\fra{2}{\l_0B\d}<\fra15\Eqa(A1.14)$$

It remains to estimate $\tilde G^{(k)}$ itself. Noting that it is the
collection of the terms that are subtracted by the above $*$
operations it is clear that it is bounded by $D B^{k-1}(k-1)!\G_0
C(\h,\d)^{-1}$ because one does not have to invert $\D$ to find it
from the bounds on the higher order terms of \equ(3.1), and one thus
saves a division by $\G_0 C(\h,\d)^{-1}$. Hence we fix
$D=E_0\G_0^{-1}C(\h,\d)$ and, with this choice, the conditions imposed
to get the above bounds, are all implied by:
$$ \fra{E_0 C(\h,\d)}{\G_0 \l_0 B\d},\
\Big(\fra{E_0 C(\h,\d)}{\G_0\l_0
\d}\Big)^2\fra1B< c'\Eqa(A1.15)$$
for some small enough $c'$.  Hence fixing an order $\NN$ the induction
works for $k\le\NN$ provided we take $\d=\x_0/4\NN$ and we find (since
$4E_0 C(\h,\d)/(\G_0\l_0\x_0)>1$, for $\h$ small enough):
$$b\,\Big(\fra{E_0 C(\h,\d) \NN}{\G_0\l_0 \x_0}\Big)^2=B,\qquad
D=\fra{E_0 C(\h,\d)}{\G_0}\Eqa(A1.16)$$
The condition on $\e$ has to be such that the map \equ(3.2) can be
defined in the domain $\DD_{\x_0/2, \r_0 e^{-2\x_0},\k_0 e^{-2\x_0}}$,
which means that:
$$|\dpr_\aa \F|\ll \x_0 \r_0,\quad
|\dpr_\AA \F|\ll \x_0,
\quad |\dpr_{p'} \F|\ll \k_0^{\fra12}\x_0,\quad
|\dpr_{q} \F|\ll \k_0^{\fra12}\x_0\Eqa(A1.17)$$
Bounding the sum $\F=\sum_{j=1}^\NN \e^j\F^{(j)}$ and $\sum_{j=1}^\NN
\e^j \tilde G^{(j)}$ via the bound \equ(A1.1), \equ(A1.16) this means
that the corresponding conditions for the existence of the canonical
maps become: $ |\e B\NN| < 1/2$, $ |\e D| \le \x_0^2 \l_0 $, which
requires $\e$ to be smaller than the minimum $\e_0$ between $(\G_0
\l_0\x_0E_0^{-1}C(\h,\d)^{-1})^2\NN^{-3} b_0$, with $b_0$ suitably small,
and $\G_0\l_0\x_0^2E_0^{-1}C(\h,\d)^{-1}b_0$:
$$|\e|<\e_0 = b_0\Big(\fra{\G_0\l_0\x_0}{E_0}\Big)^2
\fra{1}{\NN^3 C(\h,\d)^2}\Eqa(A1.18) $$
where one power of $\NN$ in the factor $\NN^{-3}$ comes from
$\NN!<\NN^\NN$ (see \equ(A1.1)) and the other two from the $\NN$ in
\equ(A1.16); and we use also $\x_0<1$ and $\G_0\l_0C(\h,\d)^{-1}
E_0^{-1}<1$ (see \equ(2.2)), so that the minimum in \equ(A1.18) is
reached in the first term.

It remains to study the remainders of order $>\NN$ in the Taylor
expansion: this can easily done in terms of $\NN$, $B$ and $D$, and a
bound $2(\e/\e_0)^\NN E_0$ is obtained. Then from the fact that
$C(\h,\d)\le c_0 \hdm \x_0^{-1}(\NN\x_0^{-1})^2\log(\NN\x_0^{-1})$,
for $\NN\le\z(\h N)^{-1}$, theorem 1 follows,
with $b'=b_0c_0^{-2}$ small enough.

\*

A {\it corollary} of theorem 1 is a {\it fast averaging} result:
namely if there are {\it no slow frequencies} (\ie if the Fourier
transform in $\aa$ contains only Fourier modes which are non-zero
multiples of a given $\nn_0=(0,\n_2)$ or more generally of
$\nn_0=(\n_1,\n_2)$ with $\n_2\ne0$, \ie if $f$ is ``{\it
monochromatic}'') the lower bound on $\D$ is $\G_0 \lis C(\h,\d)^{-1}$
with $\lis C(\h,\d)=(\hdp \d^{-2}+\d^{-1})\log \d^{-1}$. Hence if
$\NN$ is {\it fixed arbitrarily but $\h$--independent} the
condition for casting the Hamiltonian in the form \equ(4.4) is
$|\e|<O(1)$.
\*

\0{\bf References}
\*
\item{[BG] } Benettin, G., Gallavotti, G.:
{\it Stability of motions near resonances in quasi-integra\-ble
Hamiltonians systems}, Journal of Statistical Phy\-sics {\bf 44},
293-338, 1986.

\item{[CG] } Chierchia, L., Gallavotti, G.: {\it Drift and diffusion
in phase space}, Annales de l' Institut Henri Poincar\'e B {\bf 60},
1--144, 1994. See also the ``erratum'' in print  on the same journal.

\item{[E] } Eliasson, L.H.:
{\it Absolutely convergent series expansions for quasi-periodic
motions}, Mathematical Physics Electronic Journal {\bf 2},
1996 (http:// mpej.unige.ch).

\item{[G1] } Gallavotti, G.: {\it Quasi integrable mechanical
systems}, in {\it Critical phenomena, random systems, gauge theories},
Les Houches, XLIII (1984), vol. II, p. 539-- 624,
Ed. K. Osterwalder and R. Stora, North Holland, 1986.

\item{[G2] } Gallavotti, G.: {\it Hamilton--Jacobi's equation and
Arnold's diffusion near invariant tori in a priori unstable
isochronous systems}, Preprint, in mp$\_$arc@math.utexas.edu, \#97-555.

\item{[G3] } Gallavotti, G.:
{\it Twistless KAM tori, quasi flat homoclinic intersections, and
other cancellations in the perturbation series of certain completely
integrable Hamiltonian systems. A review}, Reviews on Mathematical
Physics {\bf 6}, 343--411, 1994.

\item{[G4] } Gallavotti, G.: {\it Twistless KAM tori},
Communications in Mathematical Physics {\bf 164}, 145--156, 1994.

\item{[GGM1] } Gallavotti, G. , Gentile, G., Mastropietro, V.:
{\it Separatrix splitting for systems with three degrees of freedom},
Preprint, in mp$\_$arc@math.utexas.edu, \#97-472
with the title {\it Pendulum: separatrix splitting}.

\item{[GGM2] } Gallavotti, G. , Gentile, G., Mastropietro, V.:
Lindstedt series and existence of heteroclinic chains in
Hamiltonian three time scales systems, Preprint, 1997.

\item{[Ge1] } Gentile, G: {\it A proof of existence
of whiskered tori with
quasi flat homoclinic intersections in a class of almost integrable
Hamiltonians systems}, Forum Mathematicum {\bf 7}, 709--753, 1995.

\item{[Ge2] } Gentile, G.: {\it Whiskered tori with prefixed
frequencies and Lyapunov spectrum}, Dynamics and Stability of Systems
{\bf 10}, 269--308, 1995.

\item{[P] } Perfetti, P.:
{\it Fixed point theorems in the Arnol'd model about instability of
the action-variables in phase space}, to appear in Discrete and
Continuous Dynamical Systems, in mp$\_$arc@math.utexas.edu, \#97-478.

\*

\FINE

\ciao